\documentclass[12pt, a4paper]{article}
\usepackage {amsfonts}
\usepackage [english]{babel}
\usepackage [utf8]{inputenc}
\usepackage [T1]{fontenc}
\usepackage{setspace}
\usepackage {graphicx}
\usepackage {rotate}
\usepackage{authblk}
\usepackage{anysize}
\marginsize{4cm}{2.25cm}{2.2cm}{2.2cm}

\title{Trust in foreseeing neighbours - a novel threshold model of financial market}
\author{Jan A. Lipski\thanks{j.lipski@student.uw.edu.pl} }
\author{Ryszard Kutner}
\affil{Faculty of Physics, University of Warsaw \\ Hoża 69, PL-00681 Warsaw, Poland}

\renewcommand{\thefigure}{\arabic{figure}}

\begin{document}
\maketitle
\begin{abstract}
The three-state agent-based 2D model of financial markets in the version proposed by Giulia Iori in 2002 has been 
herein extended. We have introduced the increase of herding behaviour by modelling the altering trust of an agent 
in his nearest neighbours. The trust increases if the neighbour has foreseen the price change correctly and the 
trust decreases in the opposite case. Our version only slightly increases the number of parameters present in the 
Iori model. This version well reproduces the main stylized facts observed on financial markets. That is, it reproduces 
log-returns clustering, fat-tail log-returns distribution and power-law decay in time of the volatility autocorrelation
function. 
\end{abstract}

\begin{doublespace}
\begin{section}{Introduction}
The modern study of financial markets has discovered a huge number of phenomena violating the Brownian stochastic 
dynamics of financial markets. Three major stylized facts observed in the market are exceptionally intriguing: 
volatility clustering, fat-tail log-return distribution and power-law decay in time of the volatility autocorrelation 
function. The promising approach to reproduce these phenomena runs through heteroagent-based models where decisions 
of individual agents somehow determine the price dynamics. Various versions of such models were thoroughly discussed 
in literature [1-7]. Our version well describes the above mentioned stylized facts by introducing a characteristic 
emotion to the Iori model. That is, the trust in the foreseeing market agents.
\end{section} 

\begin{section}{Outline of the Iori model}
The Giulia Iori model \cite{Iori} consists of $n$ spins or agents placed in 2D square lattice sites, assuming one of 
the following values of spin states: either $+1$ as buy, $-1$ as sell or $0$ as stay passive. Each agent, 
$i = 1, 2, ..., n$, is under the influence of a local field $Y_i$ which is the sum of the forces exerted by the nearest 
neighbours and by random noise.

Thresholds play a significant role in the Iori model. Thanks to these thresholds, the temporal spin $\sigma_i$ 
is defined as follows: 
\begin{equation}
\label{spiny}
\sigma_i(t+1) = \left\{ \begin{array}{rl}
1 &\mbox{if $Y_i(t) \geq \xi_i(t)$}\\
0 &\mbox{if $-\xi_i(t) < Y_i(t) < \xi_i(t)$}\\
-1 &\mbox{if $Y_i(t) \leq -\xi_i(t)$,}
\end{array} \right.
\end{equation}
where $\xi_i(t)$ is a time-dependent positive threshold considered below and $Y_i(t)$ is a time-dependent local field:
\begin{equation}
\label{szostka}
Y_i(t) = \sum\limits_{j = 1}^4 J_{ij}(t) + \eta_i(t), 
\end{equation}
here $J_{ij}$ is a time-dependent force exerted by agent $j$ on his neighbouring agent $i$ and $\eta_i(t)$ is a 
time-dependent white noise describing the agent's emotion. This force is defined in Section \ref{sekcja}.

The threshold $\xi_i(t)$ reflects the symmetry between the probability of buying and selling. The initial threshold 
$\xi_i(0)$ is randomly drawn from the Gaussian distribution $N(0,1)$. Next, it is adjusted in successive time steps 
after each decision round, according to the rule: 
\begin{equation} \label{2}
\frac{\xi_i(t+1)}{\xi_i(t)} = \frac{P(t)}{P(t-1)}, \ t \geq 0,
\end{equation}
where $P(t)$ is a market price of a share at time $t$. For the negative time (i.e. for history before the begining 
of the simulation) the price values $P(t<0)$ are drawn from a uniform distribution.

In Equation (\ref{2}) it was tacitly assumed: (i) the proportionality of the threshold to the transaction cost and 
(ii) proportionality of this transaction cost to the share prices. Iori proved that if the threshold is either zero 
or constant in time, the stylized facts cannot be properly reproduced. 

It is assumed that agents mutually exchange information in a consultation round and they have the opportunity 
to change their opinion (or spin state) once on average. Hence, the local field can alter. The consultation round 
is the analogy to the onr Monte Carlo step per spin in Monte Carlo simulations of magnetism. The agents begin trading 
only after the field is relaxed. It is empirically proven within the Iori model that then the most of spins become 
constant, which is an analogy to thermalisation of magnetisation in physics.

Subsequently, one calculates the total supply $S(t)$ (the total number of negative spins) and demand $D(t)$ (the total 
number of positive spins).  Thus, the market maker can determine the current market price according to the following 
rule:
\begin{equation}
\label{wzornacene}
\frac{P(t+1)}{P(t)}=  \left(\frac{D(t)}{S(t)}\right) ^{\kappa(t)},  
\end{equation}
where market activity
\begin{equation}
\label{exp4}
  \kappa(t) = \alpha\frac{D(t)+S(t)}{n}
\end{equation} 
is a slowly-varying function of time and coefficient $\alpha $ was assumed as a sufficiently small calibration 
parameter. Equation (\ref{wzornacene}) describes an asymmetric reaction of the market maker to imbalance between 
supply and demand. The intensity of this reaction (measured by the value of log-return) depends linearly on the market 
activity $\kappa(t)$. Equation (\ref{wzornacene}) is consistent with observed positive correlation between absolute 
log-returns and trading volume (defined, as usual, by quantity related to $min(D(t),S(t))$). The relation between 
demand, supply and the price change is differently considered in various agent-based models \cite{Wlosi} (and refs. 
therein). However, the models have to be consistent with the classic economic law of demand and supply.

As usual, the log-return $r$ over a time period $t-t'$ is defined as a natural logarithm of ratio of the corresponding 
prices $P(t)$ and $P(t-t')$:
\begin{equation}
\label{logret}
  r(t-t') = \ln\left(\frac{P(t)}{P(t-t')}\right),  \; t' \leq t.
\end{equation}
\end{section}
In fact, mainly this definition is exploited in our paper.

\begin{section}{Our extension: trust in the foreseeing neighbours}
\begin{subsection}{Definition of the impact}\label{sekcja}
The aim of our extension is to define a more realistic impact of a given agent on his neighbours than in the Iori 
model. This impact is large if the agent has recommended to buy the asset before a price increase or to sell before 
a price decrease. That is, the impact is large if the product of the spin value at a certain time in the past and 
the log-return from that time until now is positive. Otherwise, the impact is small. Hence, the force exerted by an 
agent $j$ on his neighbour $i$ is:
\begin{equation}
\label{soins}
J_{ij}(t) = W_{ij}+\displaystyle\sum\limits_{\tau' = 
t - \tau}^{t - 1} \sigma_j(\tau') \ln\left(\frac{P(t - 1)}{P(\tau')}\right), \ \tau \geq 2.
\end{equation}
Coefficient $J_{ij}(t=0)$ is the value of the initial force allotted at the beginning of the simulation to the pair 
of investors $i,j$. It is either $1$ with a fixed probability $p$ or $0$ with complementary probability $1-p$ (as in 
Iori model) and $W_{ij}$ is  the background static influence of agent $j$ on agent $i$, randomly drawn from a uniform 
distribution. The sum over $\tau'$ represents the dynamic part of the impact containing a kind of $\tau$-step memory.

To avoid a persistent positive feedback effect (which results in directed constant price changes), a fundamental 
behaviour of agents is introduced. This fundamental behaviour is constrained by the fundamental price and two positive 
factors $a(>1)$ and $b(<1)$. If the market price is greater than the fundamental price multiply by $a$, an agent sells 
shares. Reversely, if the market price is lower by factor $b$, an agent buys shares.  
\end{subsection}

\begin{subsection}{Algorithm}\label{algo}
Our algorithm makes possible to simulate the behaviour of agents. Single agent can trade only one share at a time 
that is, he can buy shares only if he possesses enough cash or sell shares only if he has at least one share. Leverage 
and short sale is beyond the scope of our model. 

The algorithm consists of the initial and proper parts. {\bf Within the initial part} the input values of variables 
and required parameters are prepared. This part is valid for the case $t = 0$ and $-\tau \le \tau' \leq -1$, where
$\tau =1,2,\ldots \; .$ For each value of $\tau'$ (at fixed value of $\tau$) it consists of the following steps.
\begin{enumerate}
  \item The spin state, $\sigma_i(\tau')$ ($i = 1, 2, ..., n$), of each agent is drawn from a (discrete) 3-point 
  uniform distribution.
  \item The price $P(\tau')$ is drawn from unit interval, $(0,1)$, by using a uniform continuous distribution. 
  \item The fixed values $W_{ij}$ (valid for any time) are drawn from interval $(-2,2)$ by using a uniform 
  distribution.
  \item Thanks to the above given steps we can calculate $J_{ij}(t=0)$ from Equation (\ref{soins}) and subsequently 
  $Y_{i}(t=0)$ from Equation \ref{szostka}.
  \item The intial values of the thresholds $\xi_i(t=0)$ are drawn from the normal distribution $N(0,1)$.
  \item The initial values of the $\delta$-correlated noise, $\eta_i(t=0)$, are drawn from the normal distribution 
  $N(0,1)$. 
  \item Cash and shares are granted as well to every agent as to market maker (for details see Section 
  \ref{section:comment}).
\end{enumerate}

Notably, each time step is divided into two rounds: the consultation and decision ones. In the consultation round 
only the relaxation of spins occur without changing any amount of cash and shares (of each agent and market maker). 
The change of cash and shares takes place only in the decision round. This is described in details below.
  
{\bf Within the proper part}, valid for $t \geq 1$, the dynamics of the system is simulated. In this part the 
dynamics of every agent is determined by the following algorithm. 
\begin{enumerate}
\item For $t=1$ the spin state of each agent is calculated according to Equation (\ref{spiny}), as we have already 
all quantities required (i.e. those for $t=0$). Thus, for $t=1$ all decisions of agents are known.
\item Next, agent  $i$ is drawn with probability $1/n$.
\item By using the spin values of $i$'s nearest neighbours, we construct forces $J_{ij}(t=1)$ exerted on agent $i$ 
(from his nearest neighbours $j$s) according to Equation (\ref{soins}). This is possible as all required quantities 
have been calculated one step earlier.
\item Hence, the local field $Y_i(t=1)$ is calculated according to Equation (\ref{szostka}).
\item The agent's threshold $\xi_i(t=1)$ is calculated from Equation (\ref{2}) as all required quantities have been 
calculated one step earlier.
\item Finally, the spin state $\sigma_i(t=2)$ is calculated according to Equation (\ref{spiny}).
\item The steps 2-6 are repeated until the spin relaxes. This means that decisions of agents stabilise and 
the consultation round is finished. This relaxation is, in fact, observed in our simulation quite well.
\item The final spin values are the final decisions of all agents. That is, these spin values are the output of the 
consultation round preceding the decision round in which agents are trading.
\item According to their final decisions (taken at the end of the consultation round), agents put sell or buy orders 
(being only allowed to trade one share at a given time step) however, the final share prices still have to be defined.
\item The market maker determines the price according to Equation \ref{wzornacene}.
\item The agents trade, as long as they possess enough shares or cash.
\item Agents who have no sufficient amount of cash or shares, are not able to trade with other agents. Instead, they 
can trade with the market maker (at prices established in above item) if he has sufficient amount of cash and shares. 
However, if market maker has no sufficient cash and shares the algorithm stops the current decision round and goes 
to the subsequent time step. 
\item In a given decision round $t$ (the number of decision round always equals the number of time steps) the agents 
demonstrate a fundamental behaviour with probability $\pi$ under the following conditions: 
\begin{itemize}
  \item remainder $R$ from dividing the time step (or number) $t$ ($t = 1,...,L$) by a natural number $m( = K + k)$ 
  is lower than a fixed value $\rho$ (or $R=t-Ent(t/m) m <\rho $, where $Ent(x)$ means the entire part of $x$). Number 
  $K$ is a natural number fixed at the beginning of the simulation, while $k$ is also a natural number but drawn from 
  a uniform distribution. As $k$ fluctuates, it protects the agent trading against periodicity (which could be 
  present in the case of fixed $m$). Apparently, for $t\leq K$ remainder $R$ is always smaller than $\rho$ (as it 
  vanishes);
  \item the market price of share is either higher than the fundamental price multiplied by factor $a(>1)$ (then the 
  agent can sell a share for the market price) or lower than the fundamental price by factor $b(<1)$ (then the 
  agent can buy a share for the market price). Indeed, this is a fundamental behaviour of agent because he is driven
  by the relation between the market price and the fundamental one and not by the collective impact of his neighbours. 
\end{itemize}
\end{enumerate}

\end{subsection}
\begin{subsection}{Comments concerning simulations}\label{section:comment}
The values of parameters of the model have a substantial influence on the accuracy of the reconstruction of stylized 
facts.  For instance, decreasing coefficient $\alpha $ (present in Equation (\ref{exp4})) by one order of magnitude 
(here from $\alpha=0.01$  to $\alpha=0.001$), increasing the number of agents  $n$  (here from $n=10$ to $n=15$), 
and decreasing probability $\pi $ of trading according to the fundamental price  (here from $\pi =90\%$  
to $\pi =80\%$), one gets less accurate results: less visible log-returns clustering and less accurate log-returns 
distribution.

We performed our simulations, named simulation A and simulation B, which vary only by values of some parameters. 
The parameter values of simulation A are as follows:
\begin{itemize}
\item the number of agents $n = 1024$;
\item at the beginning of the simulation each agent \emph{a'priori} received 100 cash units and the same number 
of shares. Similarly, the market maker received 10240 cash and shares;
\item $L = 80 000$ decision rounds were set and the system's memory was extended until $\tau = 20$ decision rounds;
\item the agents demonstrated a fundamental behaviour (see Algorithm in Section \ref{algo} for details) if the 
remainder of dividing the number of decision round $t$ by $m = 275+k$ (where $K=275$ and discrete variable $k$ was 
drawn from domain $k = 1,\ldots ,15$ by a uniform distribution) was lower than $\rho = 20$; thus they sell shares if 
their price is $a = 1.5$ times greater than the fundamental price and buy shares if their price is $b = 0.667$ 
times lower than their fundamental price - both with probability $\pi = 70\%$;
\item in every decision round the fundamental value of shares rise by factor $\frac{1.05}{1500}$;
\item the coefficient $\alpha$ from Equation (\ref{wzornacene}) equals $0.01$.
\end{itemize}

The values of parameters driving simulation B were assumed the same as in simulation A, except the following ones:
\begin{itemize}
  \item the system's memory $\tau = 40$ decision rounds;
  \item the agents demostrated fundamental behaviour (see Algorithm in Section \ref{algo} for details) if the 
  remainder of dividing the number of decision round $t$ by $m = 200 + k$ (where $K=200$ and discrete variable $k$ 
  was drawn from domain $1,\ldots ,10$ by a uniform distribution) was lower than $\rho = 30$; then they 
  traded according to the fundamental price analogously as for simulation A (considered above) but with larger 
  probability $\pi = 90\%$.
\end{itemize}

For the purpose of comparison, we have also conducted simulations for the case where the agents' memory was extended 
only over one time i.e. over $\tau=1$. For comparison, these simulations with \emph{no agents' esteem}, are shown 
together with simulations A and B in Figure \ref{volclust}, There the long-term rising and falling of trust in the 
foreseeing agents is somehow implicitly coded.
\end{subsection}
\end{section}

\begin{section}{Main results}
\begin{subsection}{Log-returns clustering}
The highly non-Gaussian property of log-returns as defined by Equation (\ref{logret}) of log-returns time clustering 
is being studied in this section. The results given below, showing log-returns clustring, have been obtained by 
assuming 6 decision rounds as a single  trading day. This behaviour directly implies the clustering of volatility 
(defined, e.g., as the absolute value of log-returns or their square). Our results obtained in simulations A and B, 
are shown in Figure \ref{volclust} (both middle plots), were compared with the closing time daily empirical data for 
the indicies: the Warsaw Stock Index (WIG) from 1991-04-16 to 2012-01-05 (top left plot) and the Standard 
\& Poor's 500 index from 1968-01-01 to 2012-01-05 (top right plot)  \cite{stooq}.
\renewcommand{\thefigure}{\arabic{figure}}

Apparently, the log-returns clusterings obtained in simulations for variograms are sufficiently distinct, although  
not so pronounced as for the corresponding empirical variograms. All variograms are shown in Figure \ref{volclust}, 
Notably, no periodicity is observed for the variograms despite the fact that the fundamental behaviour of investors 
has some periodic component. Our results occur not to be substantially affected if the system's memory is cut to the 
minimal range, that is to $\tau = 1$ (see the plot placed at the bottom of Figure \ref{volclust}).

\end{subsection}
\begin{subsection}{Power-law decay of autocorrelation function}
The (normalized) autocorrelation function over time $t$ as a function of the time lag $\tau$ has been calculated as 
follows:
\begin{equation}\label{problem}
 C(\tau) = \frac{\left<r(t) \, r(t-\tau)\right> - \left<r\right>^2}{Var(r)},  
\end{equation}
where $t$ is the decision round number, $\tau$ is the time lag, $\left<...\right>$ is a time average (cf. Equation 
(\ref{logret})), and $Var(r)$ is the variance of the distribution of that log-return over time period $t$.

The predictions of Equation (\ref{problem}) are compared with empirical data in Figure \ref{slabe}. Despite of poor 
agreement, some short-term relaxations are well seen, in particular, for the S\&P 500 index (bottom plots).

The autocorrelation function of absolute daily log-returns reveals long-term power-law relaxation versus time:
\begin{equation}
\label{powlaw}
R(\tau) \propto \tau^{-\gamma}, \gamma > 0.
\end{equation}
The values of exponent $\gamma$ obtained from simulations well agree with the corresponding ones obtained from 
the empirical data both for WIG and S\&P 500. We found the following values of exponent $\gamma$ for the empirical 
data namely, $\gamma(WIG) = 0.546$ and $\gamma(S\&P 500) = 0.541$ as well as for simulations $\gamma(A) = 0.576$ and 
$\gamma(B) = 0.503$ (cf. Figure \ref{decay}). Exponent $\gamma$ for simulation A agrees well with the corresponding 
one for WIG data and for simulation B agrees with that for S\$P 500 data. Moreover, the exponent $\gamma$ for the 
model without long memory of agents' esteem (i.e. for $\tau =1$) equals $\gamma(\tau=1)=0.504$, which is not so far 
placed from results obtained in simulations A and B. The results shown in Figure \ref{autovol} make our model promising for 
studying subsequent stylized facts.
\end{subsection}

\begin{subsection}{Fat-tail log-returns}
Histograms for real market daily log-returns exhibit fat-tails (see Figure \ref{Fig_4}). However, the tails are thinner 
than the corresponding tails of the L\'evy distribution. Apparently, better agreement between simulations and empirical 
data has been observed for the developed market described by S\&P 500 than for the emerging Warsaw Stock Exchange. 
This stylized fact (i.e. the presence of fat-tails in daily log-returns distributions) is not reproduced if the 
system's memory is assumed as $\tau=1$. 
\end{subsection}
\end{section}
\begin{section}{Conclusions}
Our modifications of the Iori model using success based strategy, which introduced intrinsically-driven herding 
and emotional behaviour of investors, gave satisfactory results. The following stylized facts have been reproduced: 
\begin{itemize}
\item log-returns and hence volatility clustering,
\item fat-tail log-return distribution,
\item power law autocorrelation function decay vs. time  of log-returns and  also quick (short-term) decay of 
autocorrelation function of  log-returns.
\end{itemize}
It is shown that the role of a long memory of the system, that is, a growing trust in the foreseeing neighbours, 
is most pronounce in the case of reproducing fat-tail dostribution of log-returns. 

We can conlude our (as we hope) quite representative results by the general observation that our model significantly 
better fits empirical data coming from stock markets of large size than from emerging markets.

Since the local field $Y_i$ (present in Equation (\ref{szostka})) can be interpreted as the $i$th agent's utility 
function, a bridge between our model and utility function theories has been constructed. 
\end{section}
\end{doublespace}

\listoffigures
\begin{figure}[htb]
   \includegraphics[width=0.46\textwidth]{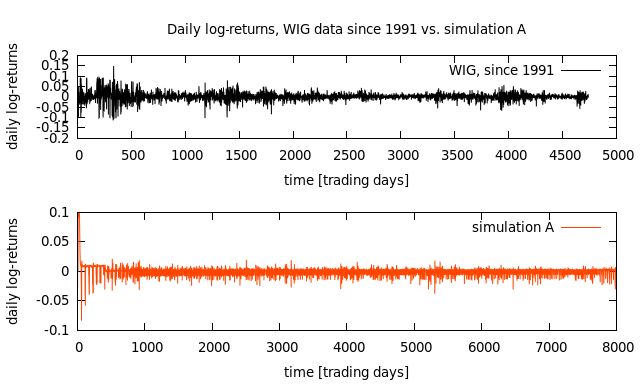}
   \includegraphics[width=0.48\textwidth]{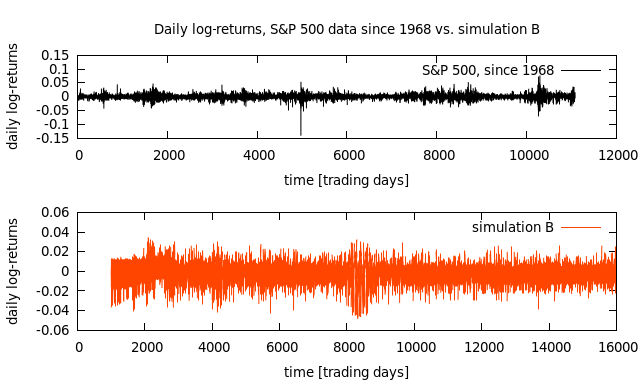}
   \begin{center}
     \includegraphics[width=0.48\textwidth]{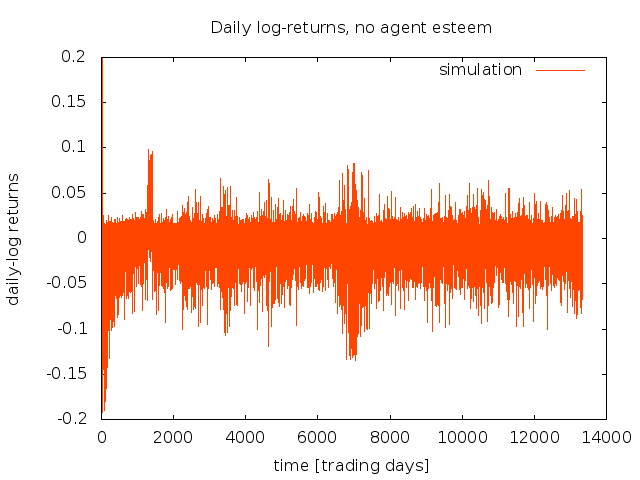}
   \end{center}
   
   \caption{Daily log-returns clustering for real market \cite{stooq} and for all our simulations. Detailed 
   descriptions of the plots are in their titles and legends.}\label{volclust}
\end{figure}
\begin{figure}[htb] 
  \includegraphics[width=0.49\textwidth]{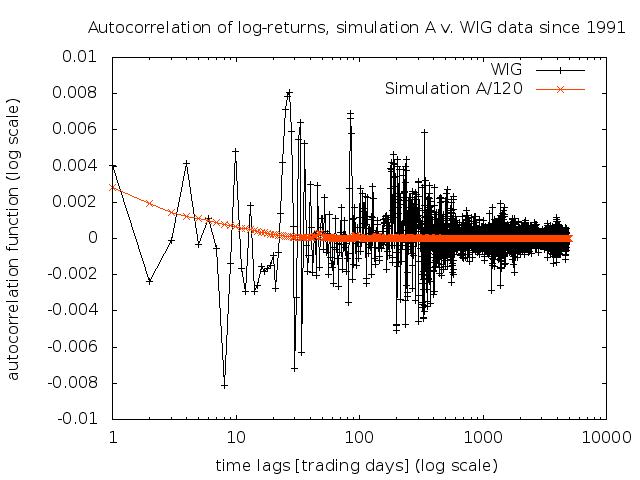} \includegraphics[width=0.49\textwidth]{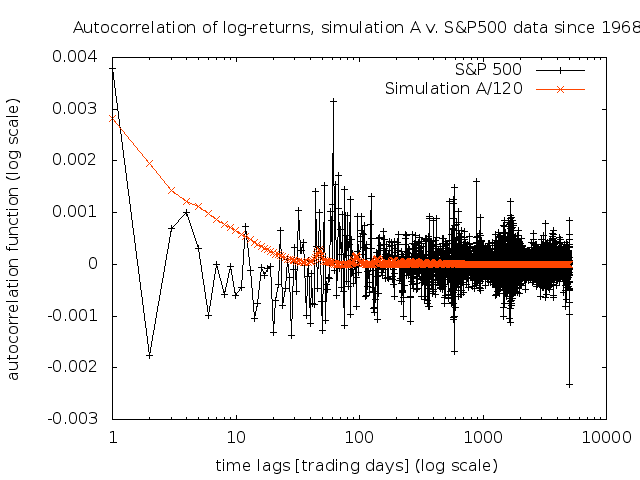}
  \includegraphics[width=0.49\textwidth]{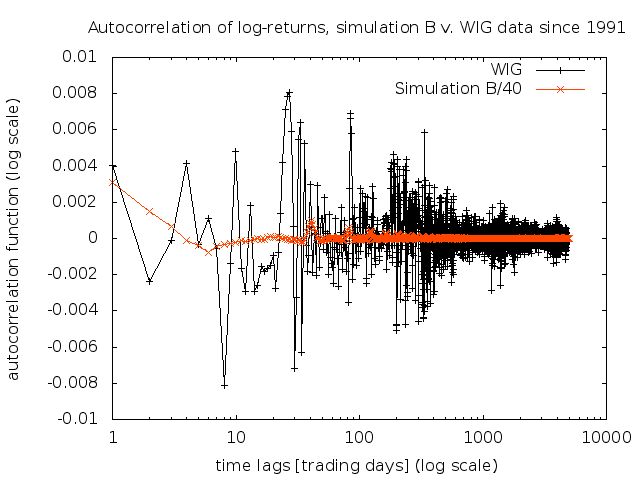} \includegraphics[width=0.49\textwidth]{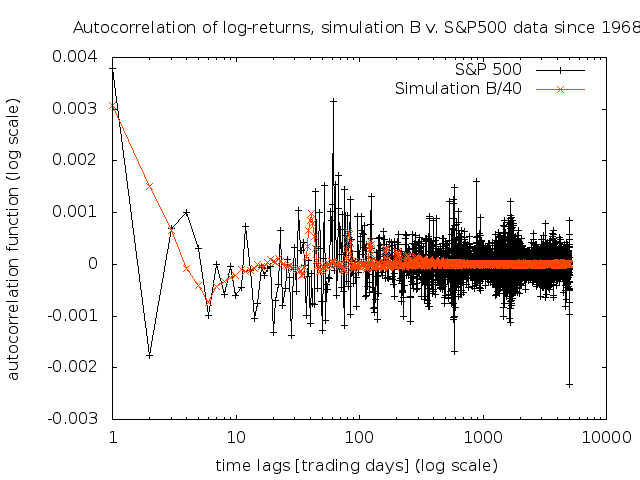}
  \caption{Autocorrelation function for daily log-returns for real markets \cite{stooq} and for our simulations 
  A and B vs. time. The autocorrelation function is less oscillating in our simulations than for the empirical data. 
  However, in both cases its short-term relaxation is well seen. Detailed descriptions 
   of the plots are in their titles and legends.} \label{slabe} 
\end{figure} 
\begin{figure}[htb]\vspace{0.05cm}\label{autovol}
  \includegraphics[width=0.49\textwidth]{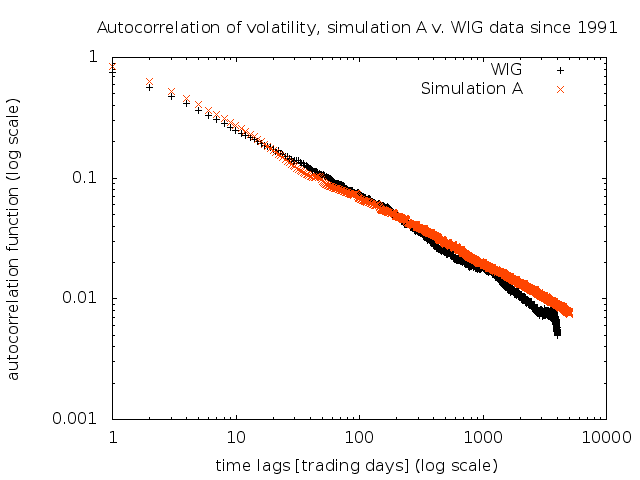} \includegraphics[width=0.49\textwidth]{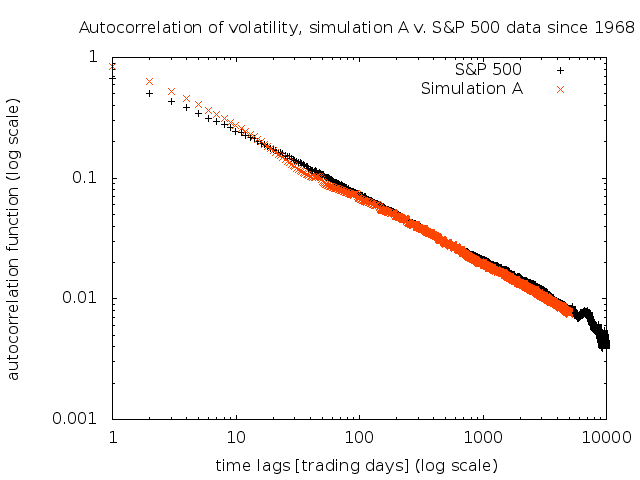}
  \includegraphics[width=0.49\textwidth]{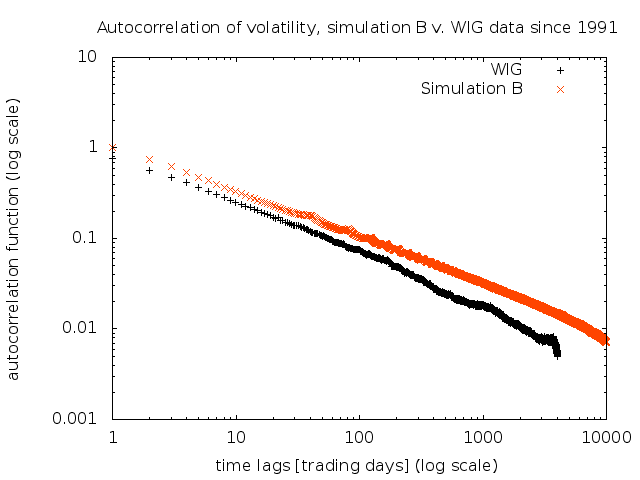} \includegraphics[width=0.49\textwidth]{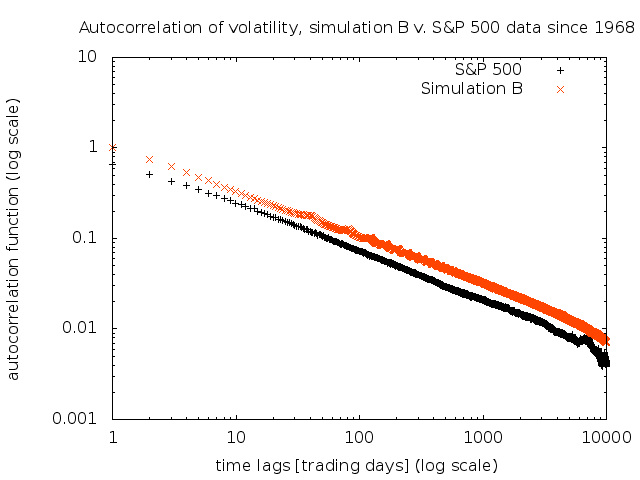}
  \begin{center}
    \includegraphics[width=0.49\textwidth]{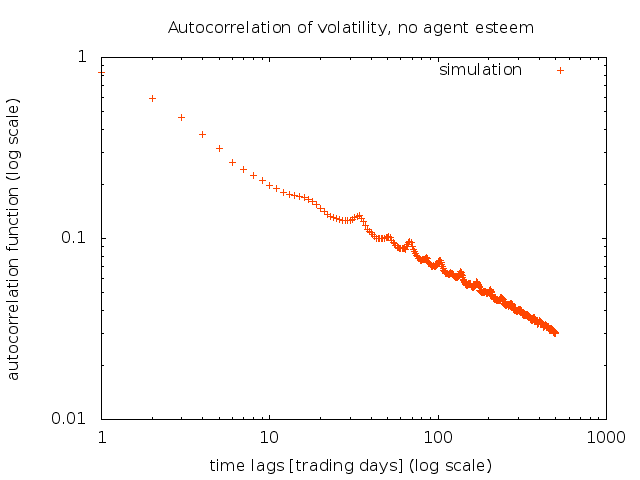}
  \end{center}
  
  \caption{The empirical (pluses) and simulational (crosses) autocorrelation functions of absolute log-returns vs. 
  time. Autocorrelation functions obtained from simulations are consistent with the corresponding ones for market data 
  \cite{stooq} for both indicies. For comparison, simulations with \emph{no agents' esteem} are also shown. 
  The power-law relaxations of all autocorrelation functions are well seen over several decades. Detailed descriptions 
  of the plots are in their titles and legends.} \label{decay} \medskip
\end{figure} 
  \begin{figure}[htb] 
\includegraphics[width=0.49\textwidth]{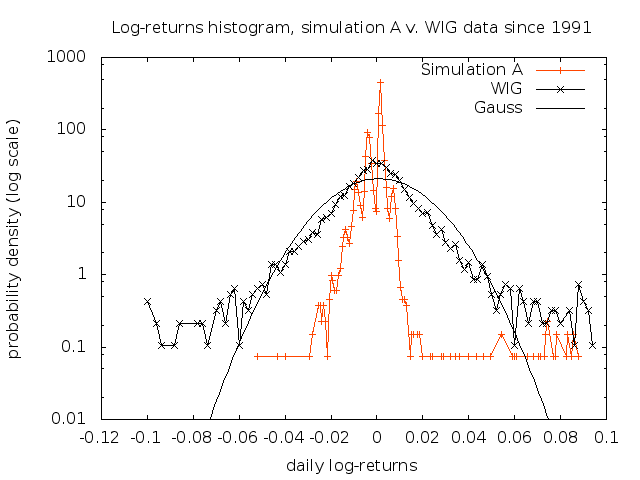} \includegraphics[width=0.49\textwidth]{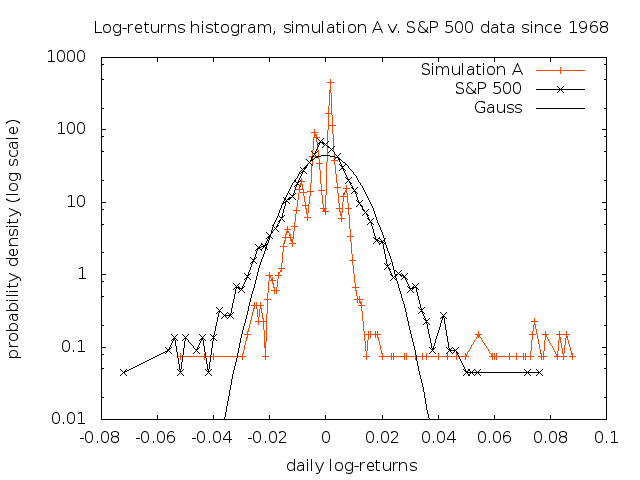} 
\includegraphics[width=0.49\textwidth]{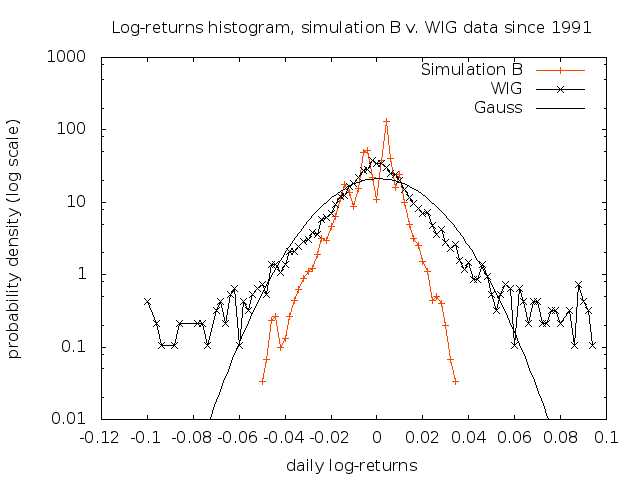} \includegraphics[width=0.49\textwidth]{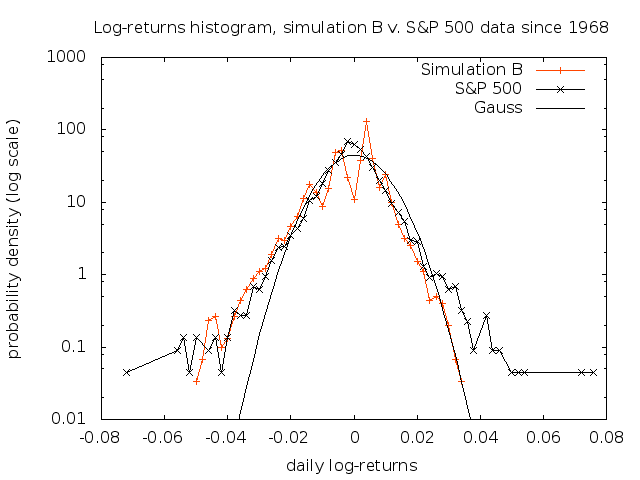}
\begin{center}
  \includegraphics[width=0.49\textwidth]{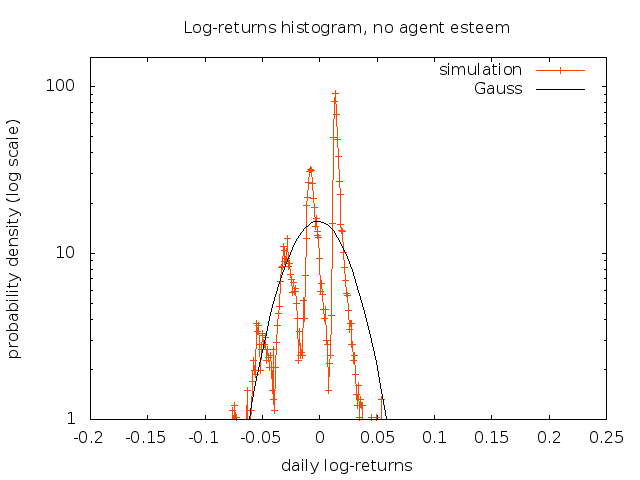}
\end{center}

\caption{Histograms of daily log-returns both for market data \cite{stooq} and obtained from simulations in the 
semi-logarithmic plots. The obtained results show the non-Gaussian properties of the price process, which is 
satisfactorily described by our model, in particular, for the stock market of a large capitalisation. 
Detailed descriptions of the plots are in their titles and legends.} \label{Fig_4}
 \end{figure} 
\end{document}